\documentclass[onecolumn,showpacs,superscriptaddress,pre]{revtex4}

\usepackage{graphicx}
\usepackage{amsmath,amssymb}

\begin{document}

\title{Comb model with slow and ultraslow diffusion}

\author{Trifce Sandev}
\affiliation{Max Planck Institute for the Physics of Complex Systems, N\"{o}thnitzer
Strasse 38, 01187 Dresden, Germany}
\affiliation{Radiation Safety Directorate, Partizanski odredi 143, P.O. Box 22, 1020
Skopje, Macedonia} 
\author{Alexander Iomin}
\affiliation{Department of Physics, Technion, Haifa 32000, Israel}
\author{Holger Kantz}
\affiliation{Max Planck Institute for the Physics of Complex Systems, N\"{o}thnitzer
Strasse 38, 01187 Dresden, Germany}
\author{Ralf Metzler}
\affiliation{Institute for Physics and Astronomy, University of Potsdam, D-14776
Potsdam-Golm, Germany}
\affiliation{Department of Physics, Tampere University of Technology, FI-33101
Tampere, Finland}
\author{Aleksei Chechkin}
\email{chechkin@pks.mpg.de}
\affiliation{Max Planck Institute for the Physics of Complex Systems, N\"{o}thnitzer
Strasse 38, 01187 Dresden, Germany}
\affiliation{Akhiezer Institute for Theoretical Physics, Kharkov 61108, Ukraine}

\date{\today}

\begin{abstract}
We consider a generalised diffusion equation in two dimensions for modeling
diffusion on a comb-like structures. We analyse the probability
distribution functions and we derive the mean squared displacement in
$x$ and $y$ directions. Different forms of the memory kernels
(Dirac delta, power-law, and distributed order) are considered. It is shown
that anomalous diffusion may occur along both $x$ and $y$
directions. Ultraslow diffusion and some more general diffusive processes
are observed as well. We give the corresponding continuous time random walk
model for the considered two dimensional diffusion-like equation 
on a comb, and we derive the
probability distribution functions which subordinate the process governed
by this equation to the Wiener process.
\end{abstract}

\maketitle

\section{Introduction}

Anomalous diffusion is typically characterized by the power-law time dependence
\begin{equation}
\label{msd}
\langle x^2(t)\rangle\simeq K_{\alpha}t^{\alpha}
\end{equation}
of the mean squared displacement (MSD) \cite{bouchaud, report,pccp}. 
Here $K_{\alpha}$ denotes the generalized
diffusion coefficient. Based on the value of the anomalous diffusion exponent
$\alpha$ we distinguish subdiffusion ($0<\alpha<1$) and superdiffusion ($\alpha>1$)
. The special value $\alpha=1$ describes normal diffusion, while
ballistic, completely directed motion, corresponds to $\alpha=2$. Subdiffusion, for
instance, is routinely measured for the charge carrier motion in amorphous
semiconductors \cite{scher}, in the cytoplasm of living biological cells \cite{lene,
golding}, or in artificially crowded liquids \cite{szymanski,lene1}. Superdiffusion
is also observed in live cells due to active motion \cite{christine,elbaum}. Large
scale simulations in three \cite{aljaz,mcguffee} and two dimensional systems
\cite{kneller,jae_prl} also exhibit (transient) anomalous diffusion. In what follows
we will also consider an ultraslow, logarithmic growth
\begin{equation}
\langle x^2(t)\rangle\simeq\log^{\gamma}t
\end{equation}
of the MSD, on which we comment further below.

Anomalous diffusive transport has been investigated in low dimensional percolation clusters and comb-like structures by introducing comb models \cite{white,weiss,havlin,havlin2,arkhincheev}. The corresponding dynamic equation for the description of the continuous comb has been introduced in Ref.~\cite{arkhincheev}
and extensively studied \cite{LZ,arkhincheev1,bi2004,ib2005,da silva,dvorkon09}.
In this approach the effective comb model equation reads
\begin{align}\label{diffusion eq on a comb}
\frac{\partial}{\partial t}P(x,y,t)=\mathcal{D}_{x}\delta(y)\frac{\partial^{2}}{\partial x^{2}}P(x,y,t) +\mathcal{D}_{y}\frac{\partial^{2}}{\partial y^{2}}P(x,y,t),
\end{align}
where the initial condition is
\begin{eqnarray}\label{initial condition}
P(x,y,0)=\delta(x)\delta(y),
\end{eqnarray}
and the boundary conditions for $P(x,y,t)$ and $\frac{\partial}{\partial q}P(x,y,t)$, $q=\{x,y\}$ are set to zero at infinity, $x=\pm\infty$, $y=\pm\infty$. Here $P(x,y,t)$ is the probability density function (PDF) to find the test particle at position $(x,y)$ at time $t$. Here, $x$ measures the direction along the backbone of the comb, while $y$ is the distance along the teeth away from the backbone. This analytical form of the model is suggested by  heuristic arguments on inhomogeneous two-dimensional diffusion, where the diagonal components of a diffusion tensor, $\mathcal{D}_{x}\delta(y)$ and $\mathcal{D}_{y}$  are the diffusion coefficients in the $x$ and $y$ directions, correspondingly.
In this effective formulation of the comb model the individual teeth of the comb
are considered smeared out and averaged along the $x$ direction.

The comb model can be described in the framework of the continuous time random
walk (CTRW) theory \cite{report}, where the particle moving
along the backbone can be trapped by diffusion along the teeth during a
waiting time. The waiting time PDF for this process in the Laplace
space is given by $\psi(s)=\frac{1}{1+s^{1/2}}$, where $\psi(s)=\mathcal{L}\left[\psi(t)\right]$ is the Laplace
transform of $\psi(t)$, meaning that the
waiting time PDF in the long time limit is of power-law form
$\psi(t)\simeq t^{-3/2}$. The distribution of jump lengths
$\lambda(x)$ is of Gaussian form with variance $\sigma^{2}$,
$\lambda(k)\simeq1-\frac{1}{2}\sigma^{2}k^{2}$,
where $\lambda(k)=\mathcal{F}\left[\lambda(x)\right]$ is the
Fourier transform of $\lambda(x)$ (see Section 2). Thus, the MSD along the backbone is of the form
$\left\langle x^{2}(t)\right\rangle\simeq t^{1/2}$. This can be concluded following the procedure in Section 2 (see also \cite{bi2004}). From this approach one also finds that the PDF
$p_{1}(x,t)=\int_{-\infty}^{\infty}dyP(x,y,t)$ satisfies the following time fractional diffusion equation
\begin{align}
\frac{\partial^{1/2}}{\partial t^{1/2}}p_{1}(x,t)=\frac{\mathcal{D}_{x}}{2\sqrt{\mathcal{D}_{y}}}\frac{\partial^{2}}{\partial x^{2}}p_{1}(x,t),
\end{align}
where
\begin{align}\label{Caputo_derivative}
\frac{\partial^{\alpha}}{\partial t^{\alpha}}f(t)=\frac{1}{\Gamma(1-\alpha)}\int_{0}^{t}dt'\frac{\frac{d}{dt'}f(t')}{(t-t')^{\alpha}}
\end{align}
is the Caputo fractional derivative \cite{podlubny} and  $0<\alpha<1$. The diffusion along the teeth is normal which can be directly concluded by integration of Eq.(\ref{diffusion eq on a comb}) over $x$, i.e., $p_{2}(y,t)=\int_{-\infty}^{\infty}dxP(x,y,t)$, and taking the boundary conditions $P(\pm\infty,y,t)=0$ and $\frac{\partial}{\partial x}P(\pm\infty,y,t)=0$. Thus, one obtains the diffusion equation for Brownian motion, i.e., $\frac{\partial}{\partial t}p_{2}(y,t)=\mathcal{D}_{y}\frac{\partial^{2}}{\partial y^{2}}p_{2}(y,t)$.

There exist different generalizations of the comb model (\ref{diffusion eq on a comb}). For example, Mendez and Iomin \cite{mendez,iomin} recently considered the following fractional comb model:
\begin{align}\label{frac diffusion eq on a comb}
\frac{\partial^{\alpha}}{\partial t^{\alpha}}P(x,y,t)=\mathcal{D}_{x}\delta(y){_{t}}I_{0+}^{1-\alpha}
\frac{\partial^{2}}{\partial x^{2}}P(x,y,t)+\mathcal{D}_{y}\frac{\partial^{2}}{\partial y^{2}}P(x,y,t),
\end{align}
where
\begin{align}\label{RLintegral}
{_{RL}I_{t}^{\alpha}}f(t)=\frac{1}{\Gamma(\alpha)}\int_{0}^{t}dt'(t-t')^{\alpha-1}f(t'),
\end{align}
is the Riemann-Liouville (R-L) fractional integral. A comb-like model of the form (\ref{frac diffusion eq on a comb})was used to describe anomalous diffusion in spiny dendrites, where the MSD along the $x$ direction has the power-law dependence $\left\langle x^{2}(t)\right\rangle\simeq t^{1-\frac{\alpha}{2}}$ on time \cite{mendez,iomin}. Similar comb-like models were used in different contexts for describing subdiffusion on a fractal comb \cite{iomin2}, the mechanism of superdiffusion of ultra-cold atoms in a one dimensional polarization optical lattice \cite{iomin3}, to describe diffusion processes on a backbone structure in presence of a drift term \cite{lenzi}, to model anomalous transport in low-dimensional composites \cite{baklanov} and electron transport in disordered nanostructured semiconductors \cite{sibatov}, or developing an effective comb-shaped configuration of antennas \cite{ieee}. A comb model approach to study and simulate
flows in cardiovascular and ventilatory systems, especially for multiscale biomathematical models related to virtual physiology is given in \cite{Thiriet}. Furthermore, different random walk models were used to describe diffusion on a comb, to mention \cite{weiss,havlin2,havlin3,barkai2013,zaburdaev,LZ} but a few. Here we note that, as shown in \cite{lenzi njp}, a generalization of the comb model with correlations in the $y$ direction affects the diffusive behavior in the $x$ direction in a non-trivial fashion, resulting in a quite rich diffusive scenario of anomalous transport in the $x$ direction. Recently, it was shown
\cite{sandev iomin kantz} that if we consider discrete model with a finite number of backbones, which means that the diffusion along the $x$ direction may occur on
many backbones, located at $y=l_j$, $j=1,2,\dots,N$, $0<l_1<l_2<\dots<l_N$, the transport exponent does not change, i.e., it is equal to $1/2$. Contrary to this, in case of infinite number of backbones, which are at positions $y$ which belong to the fractal set $S_\nu$ with fractal dimension $0<\nu<1$, the transport exponent depends on the fractal dimension $\nu$. We stop to mention that the diffusion on a comb in the
$x$ direction effectively is of the continuous time random walk class and thus
weakly non-ergodic and ageing \cite{he,johannes}, while the random motion on a
fractal such as a critical percolation cluster is ergodic \cite{meroz,yousof}.

This paper is organized as follows. In Section II we introduce our generalized comb-like model
(\ref{diffusion like eq on a comb}) that has two different memory kernels. The PDF and MSD are derived and we find the corresponding CTRW models. In Section III we analyze the subordination of a process governed by Eq.(\ref{diffusion like eq on a comb}) to the Wiener process. The conditions that should be satisfied by the memory kernels in order that the PDFs are non-negative functions are discussed. Analytical results for different forms of the memory kernels are given in Section IV. It is shown that anomalous and ultraslow diffusion may occur in both directions, $x$ and $y$. A summary is provided in Section V.

\section{Generalized comb model}

Here we introduce and analyze the following generalized comb-like model
\begin{align}\label{diffusion like eq on a comb}
\int_{0}^{t}dt'\,\gamma(t-t')\frac{\partial}{\partial t'}P(x,y,t')
=\mathcal{D}_{x}\delta(y)\int_{0}^{t}dt'\,\eta(t-t')\frac{\partial^{2}}{\partial x^{2}}P(x,y,t') +\mathcal{D}_{y}\frac{\partial^{2}}{\partial y^{2}}P(x,y,t),
\end{align}
with the same initial and boundary conditions as in Eq.(\ref{diffusion eq on a comb}). In what follows we use dimensionless variables without loss of generality. Thus, we set $\mathcal{D}_{x}=\mathcal{D}_{y}=1$. Here $\gamma(t)$ and $\eta(t)$ are integrable non-negative memory kernels which approach zero in the long time limit (see next section for more details). The memory kernel $\gamma(t)$ represents the memory effects of the system, which means that the particles moving along the $y$ direction, i.e., along the teeth, may also be trapped, thus the diffusion along the $y$ direction may also be anomalous
\cite{sandev draft}. This can be directly concluded by integration of
Eq.(\ref{diffusion like eq on a comb}) over $x$, and taking the boundary conditions which are of 
the form $P(\pm\infty,y,t)=0$ and $\frac{\partial}{\partial x}P(\pm\infty,y,t)=0$, from where we obtain the generalized diffusion equation recently considered in \cite{sandev draft}
\begin{align}\label{p1 diff eq}
\int_{0}^{t}dt'\,\gamma(t-t')\frac{\partial}{\partial t'}p_{2}(y,t')=
\frac{\partial^{2}}{\partial y^{2}}p_{2}(y,t),
\end{align}
where $p_{2}(y,t)=\int_{-\infty}^{\infty}dxP(x,y,t)$. It is shown that this model represents a useful tool to describe anomalous diffusion, in particular, ultraslow logarithmic diffusion, and that it corresponds to a CTRW model with a Gaussian type jump length PDF $\lambda(y)$, and waiting time PDF given in the Laplace space by $\psi(s)=\frac{1}{1+s\gamma(s)}$. Going back to
Eq.(\ref{diffusion like eq on a comb}), the case $\gamma(t)=\eta(t)=\delta(t)$ yields the diffusion equation (\ref{diffusion eq on a comb}). The memory kernel $\eta(t)$ was introduced in \cite{mendez} to compensate the anomalous transport in the backbone. Thus, here, we call the memory kernel $\eta(t)$ as a generalized 
compensate kernel.

\subsection{Solution in Fourier-Laplace space}

In order to solve Eq.(\ref{diffusion like eq on a comb}) we use the Laplace
\begin{align}
\mathcal{L}\left[f(t)\right]=\int_{0}^{\infty}dtf(t)e^{-st}=F(s),
\end{align}
and Fourier transform,
\begin{align}
\mathcal{F}\left[f(x)\right]=\int_{-\infty}^{\infty}dxf(x)e^{\imath kx}=F(k),
\end{align}
and consequently the inverse Laplace $\mathcal{L}^{-1}\left[F(s)\right]=f(t)$, and Fourier transform $\mathcal{F}^{-1}\left[F(k)\right]=\frac{1}{2\pi}\int_{-\infty}^{\infty}dkF(k)e^{-\imath kx}=f(x)$. In what follows the transformed quantities are denoted by explicit dependence on the respective variables. By Laplace transform we find
\begin{align}\label{diffusion like eq on a comb Laplace transform}
\gamma(s)\left[sP(x,y,s)-P(x,y,t=0)\right]
=\delta(y)\eta(s)\frac{\partial^{2}}{\partial x^{2}}P(x,y,s)+\frac{\partial^{2}}{\partial y^{2}}P(x,y,s),
\end{align}
where $P(k,y,t=0)$ is the initial condition to be specified below. By Fourier
transform with respect to both space variables $x$ and $y$ it follows that
\begin{align}\label{diffusion like eq on a comb Laplace-Fourier transform}
\gamma(s)\left[sP(k_{x},k_{y},s)-
P(k_{x},k_{y},t=0)\right]=-k_{x}^{2}\eta(s)P(k_{x},y=0,s)-
k_{y}^{2}P(k_{x},k_{y},s).
\end{align}
Therefore, PDF in the Fourier-Laplace space is given by
\begin{align}\label{diffusion like eq on a comb PDF in Laplace-Fourier space}
P(k_{x},k_{y},s)=\frac{\gamma(s)P(k_{x},k_{y},t=0)-
k_{x}^{2}\eta(s)
P(k_{x},y=0,s)}{s\gamma(s)+k_{y}^{2}}.
\end{align}
The initial condition is given by (\ref{initial condition}), thus by inverse Fourier transform in respect to $k_{y}$ we find
\begin{align}\label{diffusion like eq on a comb PDF in Laplace-Fourier space InverseFy}
P(k_{x},y,s)=\frac{1}{s}\sqrt{\frac{s\gamma(s)}{4}}e^{-\sqrt{s\gamma(s)}|y|}
-\frac{1}{s}\sqrt{\frac{s\gamma(s)}{4}}\frac{\eta(s)k_{x}^{2}}{\gamma(s)}P(k_{x},y=0,s)e^{-\sqrt{s\gamma(s)}|y|}.
\end{align}
If we substitute $y=0$ we obtain
\begin{align}\label{diffusion like eq on a comb PDF in Laplace-Fourier space InverseFy0}
P(k_{x},y=0,s)=\frac{\frac{1}{s}\sqrt{\frac{s\gamma(s)}{4}}}{1+\frac{1}{s}\sqrt{\frac{s\gamma(s)}{4}}\frac{\eta(s)}{\gamma(s)}k_{x}^{2}},
\end{align}
from where it follows
\begin{align}\label{diffusion like eq on a comb PDF in Laplace-Fourier space InverseFy exact}
P(k_{x},y,s)=\frac{\frac{1}{s}\sqrt{\frac{s\gamma(s)}{4}}}{1+\frac{1}{s}\sqrt{\frac{s\gamma(s)}{4}}\frac{\eta(s)}{\gamma(s)}k_{x}^{2}}
e^{-\sqrt{s\gamma(s)}|y|},
\end{align}
and
\begin{align}\label{diffusion like eq on a comb PDF in Laplace-Fourier space exact}
P(k_{x},k_{y},s)=
\frac{s\gamma(s)\xi(s)}{\left(s\gamma(s)+k_{y}^{2}\right)\left(s\xi(s)+\frac{1}{2}k_{x}^{2}\right)},
\end{align}
where
\begin{align}\label{xi(s)} \xi(s)=\frac{1}{\eta(s)}\sqrt{\frac{\gamma(s)}{s}}.
\end{align}
Relation (\ref{diffusion like eq on a comb PDF in Laplace-Fourier space exact}) for $k_{y}=0$ yields
\begin{align}\label{diffusion like eq on a comb PDF in Laplace-Fourier space exact ky0}
P(k_{x},k_{y}=0,s)=
\frac{\xi(s)}{s\xi(s)+\frac{1}{2}k_{x}^{2}},
\end{align}
and for $k_{x}=0$
\begin{align}\label{diffusion like eq on a comb PDF in Laplace-Fourier space exact kx0}
P(k_{x}=0,k_{y},s)=
\frac{\gamma(s)}{s\gamma(s)+k_{y}^{2}},
\end{align}
corresponding to the spatial averages in $y$ and $x$ directions, respectively.

\subsection{PDF and MSD in the $x$ direction}

Let us now analyze the PDF $p_{1}(x,t)=\int_{-\infty}^{\infty}dyP(x,y,t)$. Its 
Fourier-Laplace transform reads $p_{1}(k_{x},s)=P(k_{x},k_{y}=0,s)$, i.e.
following Eq.(\ref{diffusion like eq on a comb PDF in Laplace-Fourier space exact ky0}),
\begin{align}\label{p1 general}
p_{1}(k_{x},s)=\frac{\xi(s)}{s\xi(s)+\frac{1}{2}k_{x}^{2}}.
\end{align}
From the results given in \cite{sandev draft} we conclude that the PDF $p_{1}(k_{x},s)$ corresponds to the one obtained from the CTRW theory for a process with Gaussian jump length PDF and waiting time PDF whose Laplace transform is
 $\psi_{1}(s)=\frac{1}{1+s\xi(s)}$. The normalization condition requires
$\lim_{s\rightarrow0}s\xi(s)=\lim_{t\rightarrow\infty}\xi(t)=0$. Thus, 
the memory kernel (\ref{xi(s)}) is of the form that should satisfy $\lim_{s\rightarrow0}\frac{\sqrt{s\gamma(s)}}{\eta(s)}=0$. From Eq.(\ref{p1 general}) it follows that
\begin{align}
\label{FL diffusion}
\xi(s)\left[sp_{1}(k_{x},s)-1\right]=-\frac{1}{2}k_{x}^{2}p_{1}(k_{x},s),
\end{align}
from where one finds \cite{sandev draft} that the PDF $p_{1}(x,t)$ satisfies the following generalized diffusion equation
\begin{align}
\label{diffusion-like eq memory}
\int_{0}^{t}dt'\,\xi(t-t')\frac{\partial}{\partial t'}p_{1}(x,t')=\frac{1}{2}
\frac{\partial^2}{\partial x^2}p_{1}(x,t),
\end{align}
with initial condition $p_{1}(x,0)=\delta(x)$.

The MSD therefore yields in the form \cite{sandev draft}
\begin{align}\label{diffusion like eq on a comb MSD}
\left\langle x^{2}(t)\right\rangle=\mathcal{L}^{-1}\left.\left[-\frac{\partial^{2}}{\partial k_{x}^{2}}p_{1}(k_{x},s)\right]\right|_{k_{x}=0}
=\mathcal{L}^{-1}\left[\frac{1}{s^{2}\xi(s)}\right]=\mathcal{L}^{-1}\left[\frac{1}{s}\frac{\eta(s)}{\sqrt{s\gamma(s)}}\right].
\end{align}
From this relation we directly obtain the MSD for the simple comb-like model (\ref{diffusion eq on a comb}).

\subsection{PDF and MSD in the $y$ direction}

Next we analyze the PDF $p_{2}(y,t)=\int_{-\infty}^{\infty}dxP(x,y,t)$, 
for which we find that $p_{2}(k_{y},s)=P(k_{x}=0,k_{y},s)$, i.e,
following Eq.(\ref{diffusion like eq on a comb PDF in Laplace-Fourier space exact kx0})
\begin{align}\label{p2 general}
p_{2}(k_{y},s)=\frac{\gamma(s)}{s\gamma(s)+k_{y}^{2}}.
\end{align}
According to the results given in \cite{sandev draft}, the PDF $p_{2}(y,t)$ corresponds to the one obtained from the CTRW theory for a process with Gaussian type jump length PDF and waiting time PDF given by its Laplace trasnform as
$\psi_{2}(s)=\frac{1}{1+s\gamma(s)}$. The normalization condition requires
$\lim_{\rightarrow0}s\gamma(s)=\lim_{t\rightarrow\infty}\gamma(t)=0$ as well. We rewrite (\ref{p2 general}) as
\begin{align}
\label{FL diffusion2}
\gamma(s)\left[sp_{2}(k_{y},s)-1\right]=-k_{y}^{2}p_{2}(k_{y},s),
\end{align}
from where by inverse Fourier-Laplace transform it is obtained that the PDF $p_{2}(x,t)$ satisfies the following generalized diffusion equation
\begin{align}
\label{diffusion-like eq memory2}
\int_{0}^{t}dt'\,\gamma(t-t')\frac{\partial}{\partial t'}p_{2}(y,t')=
\frac{\partial^2}{\partial y^2}p_{2}(y,t),
\end{align}
with initial condition $p_{2}(y,0)=\delta(y)$.

Thus, for the MSD along $y$ direction one finds \cite{sandev draft}
\begin{align}\label{MSD y direction}
\left\langle
y^{2}(t)\right\rangle=\mathcal{L}^{-1}\left.\left[-\frac{\partial^{2}}{\partial k_{y}^{2}}p_{2}(k_{y},s)\right]\right|_{k_{y}=0}=
2\mathcal{L}^{-1}\left[\frac{1}{s^{2}\gamma(s)}\right],
\end{align}
i.e., the MSD along the $y$ direction depends only on the memory kernel $\gamma(t)$.

\section{Subordination to the Wiener process}

Let us now find the corresponding PDFs which subordinate the diffusion processes, governed by equations (\ref{p1 general}) and (\ref{p2 general}), from time scale $t$ (physical time) to the Wiener processes on a time scale $u$ (operational time). In such a scheme the PDF $P(x,t)$ of a given random process $x(t)$ can be represented as \cite{barkai pre2001,chechkin pre2009,mmnp}
\begin{equation}
\label{wienersub}
P(x,t)=\int_0^{\infty}P_0(x,u)h(u,t)du
\end{equation}
where
\begin{equation}
\label{Wiener PDF}
P_0(x,u)=\frac{1}{\sqrt{4\pi u}}\exp\left(-\frac{x^2}{4u}\right),
\end{equation}
is the famed Gaussian PDF, i.e., the PDF of the Wiener process, and $h(u,t)$ is a PDF \emph{subordinating\/} the random process $x(t)$ to the Wiener process. Note that here we use that diffusion coefficient is equal to one ($\mathcal{D}=1$).

Relations (\ref{p1 general}) and (\ref{p2 general}) 
can be rewritten to
\begin{align}\label{PDF FL p1}
p_{1}(k_{x},s)=\int_{0}^{\infty}du\,e^{-u\frac{1}{2}k_{x}^{2}}h_{1}(u,s),
\end{align}
where
\begin{align}\label{h1(u,s)}
h_{1}(u,s)=\xi(s)e^{-us\xi(s)}=-\frac{\partial}{\partial u}\frac{1}{s}L_{1}(s,u),
\end{align}
\begin{align}\label{L1(s,u)}
L_{1}(s,u)=e^{-us\xi(s)},
\end{align}
and
\begin{align}\label{PDF FL p2}
p_{2}(k_{y},s)=\int_{0}^{\infty}du\,e^{-uk_{y}^{2}}h_{2}(u,s),
\end{align}
where
\begin{align}\label{h2(u,s)}
h_{2}(u,s)=\gamma(s)e^{-us\gamma(s)}=-\frac{\partial}{\partial u}\frac{1}{s}L_{2}(s,u),
\end{align}
\begin{align}\label{L2(s,u)}
L_{2}(s,u)=e^{-us\gamma(s)},
\end{align}
respectively. By inverse Fourier-Laplace transform of (\ref{PDF FL p1}) and (\ref{PDF FL p2}) one finds
\begin{align}\label{PDF p1 general}
p_{1}(x,t)=\int_{0}^{\infty}du\,\frac{1}{\sqrt{2\pi u}}e^{-\frac{x^{2}}{2u}}h_{1}(u,t),
\end{align}
\begin{align}\label{PDF p1 general2}
p_{2}(y,t)=\int_{0}^{\infty}du\,\frac{1}{\sqrt{4\pi u}}e^{-\frac{y^{2}}{4u}}h_{2}(u,t),
\end{align}
which means that the PDFs $h_{1}(u,t)$ and $h_{2}(u,t)$ provide subordination of the random processes governed by equations (\ref{p1 general}) and (\ref{p2 general}), respectively, to the Wiener process by using the operational time $u$. Here we note that the functions $h_{1}(u,t)$ and $h_{2}(u,t)$ are normalized in respect to $u$. This can be shown as follows. From Eq.(\ref{h1(u,s)}) we find
\begin{align}\label{h1(u,s) norm}
\int_{0}^{\infty}du\,h_{1}(u,t)=\mathcal{L}_{s}^{-1}\left[\int_{0}^{\infty}du\,\xi(s)e^{-us\xi(s)}\right]=\mathcal{L}_{s}^{-1}\left[\frac{1}{s}\right]=1.
\end{align}
Same procedure can be applied for $h_{2}(u,t)$.

By using the subordination approach one can see that the PDFs $p_{1}(x,t)$ and $p_{2}(y,t)$ are non-negative if respectively, $h_{1}(u,t)$ and $h_{2}(u,t)$ are
non-negative, therefore, it is sufficient to show that their Laplace transforms 
$h_{1}(u,s)$ and $h_{2}(u,s)$ are completely monotone functions 
in respect to $s$ (see Appendix A for definitions). In order $h_{1}(u,s)$, Eq.(\ref{h1(u,s)}), to be a completely monotone function, both functions $\xi(s)=\frac{1}{\eta(s)}\sqrt{\frac{\gamma(s)}{s}}$ and $e^{-us\xi(s)}$ should be completely monotone. The function $e^{-us\xi(s)}$ is a completely monotone if $s\xi(s)=\frac{1}{\eta(s)}\sqrt{s\gamma(s)}$ is a Bernstein function \cite{sandev draft}. Similarly, in order the function $h_{2}(u,s)$, Eq.(\ref{h2(u,s)}), to be a completely monotone function, the function $\gamma(s)$ should be completely monotone and  $s\gamma(s)$ is a Bernstein function. 
These conditions give restrictions on the possible choice of the kernels
$\gamma(t)$ and $\eta(t)$ in Eq.(\ref{diffusion like eq on a comb}).

\section{Different diffusion regimes}

Here we consider different forms of the memory kernels $\gamma(t)$ and $\eta(t)$, namely, Dirac $\delta$ memory kernel, power-law memory kernel, two power-law memory kernels, distributed order memory kernels, as well as their combinations. 

\subsection{Cases with $\eta(t)=\delta(t)$}

Firstly we consider cases with $\eta(t)=\delta(t)$. For the classical comb model (\ref{diffusion eq on a comb}), $\eta(t)=\gamma(t)=\delta(t)$, i.e., $\eta(s)=\gamma(s)=1$, and $\xi(s)=s^{-1/2}$. Since $\xi(s)=s^{-1/2}$ is completely monotone, and $s\xi(s)=s^{1/2}$ is a Bernstein function, then $p_{1}(x,t)$ is a non-negative function. From the other side, since $\gamma(s)=1$ is completely monotone, and $s\gamma(s)=s$ is a Bernstein function, it follows that $p_{2}(y,t)$ is non-negative function. The solution of Eq.(\ref{diffusion eq on a comb}) reads
\begin{equation}
\label{p1 delta delta}
p_{1}(x,t)=\mathcal{F}^{-1}\left[E_{1/2}\left(-\frac{1}{2}k_{x}^{2}t^{1/2}\right)
\right]=\frac{1}{2|x|}H_{1,1}^{1,0}\left[\left.\frac{|x|}{\left(t^{1/2}/2\right)^{
1/2}}\right|\begin{array}{l}\displaystyle\left(1,\frac{1}{4}\right)\\(1,1)
\end{array}\right],
\end{equation}
and the PDF along the $y$ direction has Gaussian form. For these memory kernels it follows from Eqs.(\ref{MSD y direction}) and (\ref{diffusion like eq on a comb MSD}) that the diffusion along the $y$ direction is normal $\left\langle
y^{2}(t)\right\rangle=2t$, and anomalous subdiffusion is in the $x$ direction $\left\langle
x^{2}(t)\right\rangle=2\frac{t^{1/2}}{\Gamma(1/2)}$.

Next we use power-law memory kernel $\gamma(t)=\frac{t^{-\alpha}}{\Gamma(1-\alpha)}$, i.e., $\gamma(s)=s^{\alpha-1}$, where $0<\alpha<1$. Since $\gamma(s)=s^{\alpha-1}$ is completely monotone for $0<\alpha<1$, and $s\gamma(s)=s^{\alpha-1/2}$ is a Bernstein function for $0<\alpha/2<1$, then it follows that $p_{1}(x,t)$ is a non-negative function for $0<\alpha<1$. Thus, the PDF along the $x$ direction, by using relations (\ref{ML three}), (\ref{ML three Laplace}) and (\ref{cosine H}), becomes
\begin{equation}\label{p1 power delta}
p_{1}(x,t)=\mathcal{F}^{-1}\left[E_{\alpha/2}\left(-\frac{1}{2}k_{x}^{2}t^{\alpha/2}
\right)\right]=\frac{1}{2|x|}H_{1,1}^{1,0}\left[\left.\frac{|x|}{\left(\frac{1}{2}t
^{\alpha/2}\right)^{1/2}}\right|\begin{array}{l}\displaystyle\left(1,\frac{\alpha}{
4}\right)\\(1,1)\end{array}\right],
\end{equation}
and the MSD, by employing the relation (\ref{integral of H}) grows subdiffusively,
\begin{align}\label{MSD power delta}
\left\langle x^{2}(t)\right\rangle=\frac{t^{\alpha/2}}{\Gamma\left(1+\alpha/2\right)}.
\end{align}
For the $y$ direction, in the same way, we find
\begin{align}\label{p2 power delta}
p_{2}(y,t)=\mathcal{F}^{-1}\left[E_{\alpha}\left(-k_{y}^{2}t^{\alpha}\right)\right]=\frac{1}{2|x|}H_{1,1}^{1,0}\left[\left.\frac{|y|}{t^{\alpha/2}}\right|\begin{array}{c
l}
    \displaystyle {(1,\frac{\alpha}{2})}\\
    \displaystyle {(1,1)}
  \end{array}\right],
\end{align}
and
\begin{align}\label{MSDy power delta}
\left\langle y^{2}(t)\right\rangle=2\frac{t^{\alpha}}{\Gamma\left(1+\alpha\right)},
\end{align}
i.e. subdiffusion along $y$ direction is faster than the subdiffusion along $x$ direction. Note that in the same way as in case of one fractional exponent, one can show that $p_{1}(x,t)$ is non-negative in case of the memory kernel with two fractional exponents, which we consider below.

Thus, for memory kernel of the form $\gamma(t)=C_{1}\frac{t^{-\alpha_{1}}}{\Gamma(1-\alpha_{1})}+C_{2}\frac{t^{-\alpha_{2}}}{\Gamma(1-\alpha_{2})}$, $C_{1},C_{2}>0$, $C_{1}+C_{2}=1$, $0<\alpha_{1}<\alpha_{2}<1$, for the MSD in the $x$ direction we find
\begin{align}\label{MSD two power delta}
\left\langle x^{2}(t)\right\rangle=\frac{1}{\sqrt{C_{2}}}\mathcal{L}^{-1}\left[\frac{s^{-\alpha_{1}/2-1}}{\left(s^{\alpha_{2}-\alpha_{1}}+\frac{C_{1}}{C_{2}}\right)^{1/2}}\right]
=\frac{1}{\sqrt{C_{2}}}t^{\alpha_{2}/2}E_{\alpha_{2}-\alpha_{1},\alpha_{2}/2+1}^{1/2}\left(-\frac{C_{1}}{C_{2}}t^{\alpha_{2}-\alpha_{1}}\right).
\end{align}
From here we conclude that decelerating subdiffusion appears along the $x$ direction since for the short time limit the MSD behaves as $\left\langle x^{2}(t)\right\rangle\simeq\frac{1}{\sqrt{C_{2}}}\frac{t^{\alpha_{2}/2}}{\Gamma(1+\alpha_{2}/2)}$, where we use the definition (\ref{ML three}), and in the long time limit as $\left\langle x^{2}(t)\right\rangle\simeq\frac{}{\sqrt{C_{1}}}\frac{t^{\alpha_{1}/2}}{\Gamma(1+\alpha_{1}/2)}$, where we apply asymptotic expansion formula (\ref{ML three asymptotic}). Along the $y$ direction it follows
\begin{align}\label{MSDy two power delta}
\left\langle y^{2}(t)\right\rangle=\frac{2}{C_{2}}\mathcal{L}^{_1}\left[\frac{s^{-\alpha_{1}-1}}{s^{\alpha_{2}-\alpha_{1}}+\frac{C_{1}}{C_{2}}}\right]
=\frac{2}{C_{2}}t^{\alpha_{2}}E_{\alpha_{2}-\alpha_{1},\alpha_{2}+1}\left(-\frac{C_{1}}{C_{2}}t^{\alpha_{2}-\alpha_{1}}\right),
\end{align}
i.e., decelerating subdiffusion as well, since the MSD turns from $\left\langle x^{2}(t)\right\rangle\simeq\frac{2}{C_{2}}\frac{t^{\alpha_{2}}}{\Gamma(1+\alpha_{2})}$ in the short time limit to $\left\langle x^{2}(t)\right\rangle\simeq\frac{2}{C_{2}}\frac{t^{\alpha_{1}}}{\Gamma(1+\alpha_{1})}$ in the long time limit.

Next we introduce kernels of distributed order. Thus, firstly we consider the following kernel \cite{kochubei,chechkin,chechkin2,chechkin chapter,mainardi_book}
\begin{align}\label{distibuted memory kernel gamma}
\gamma(t)=\int_{0}^{1}d\alpha\,\frac{t^{-\alpha}}{\Gamma(1-\alpha)},
\end{align}
i.e., $\gamma(s)=\frac{s-1}{s\log(s)}$, and $\eta(t)=\delta(t)$. Thus, $\xi(s)=\frac{1}{s}\sqrt{\frac{s-1}{\log(s)}}$. Such memory kernel was used in \cite{kochubei} in the theory of evolution equations with distributed order derivative, and in \cite{sandev pla} as a friction kernel in the generalized Langevin equation. The distributed order diffusion equations are introduced by Chechkin et al. \cite{chechkin,chechkin2} and they show that the corresponding MSD shows accelerating, decelerating and ultraslow diffusive behaviors. If we substitute (\ref{distibuted memory kernel gamma}) in the relations for MSDs (\ref{diffusion like eq on a comb MSD}) and (\ref{MSD y direction}), we obtain the following generalized ultraslow diffusive behaviors in the long time limit,
\begin{align}\label{MSDx distributed 1}
\left\langle x^{2}(t)\right\rangle\simeq\log^{1/2}t,
\end{align}
\begin{equation}\label{MSDy distributed 1}
\left\langle y^{2}(t)\right\rangle=2\left[\gamma+\log{t}+e^{t}\mathrm{E}_{1}(t)\right]\simeq2\log{t},
\end{equation}
where $\gamma=0.577216$ is the Euler-Mascheroni (or Euler's) constant,
and $\mathrm{E}_{1}(t)$ is the exponential integral function \cite{erdelyi},
\begin{align} \mathrm{E}_{1}(t)=-\mathrm{Ei}(-t)=\int_{t}^{\infty}dx\,\frac{e^{-x}}{x}.
\end{align}
Here we note that for calculation of MSD (\ref{MSDx distributed 1}) along the $x$ direction we apply the Tauberian theorem for slowly varying functions (see Appendix C). Thus, ultraslow diffusion occurs along both $x$ and $y$ directions.

More general distributed order memory kernel is \cite{chechkin}
\begin{equation}\label{distibuted memory kernel gamma nu}
\gamma(t)=\int_{0}^{1}d\alpha\,\nu\alpha^{\nu-1}\frac{t^{-\alpha}}{\Gamma(1-\alpha)},
\end{equation}
where $\nu>0$, and we use $\eta(t)=\delta(t)$. In the long time limit
$\gamma(s)\simeq\frac{\Gamma(1+\nu)}{s\log^{\nu}\frac{1}{s}}$, and for 
the MSDs we obtain
\begin{equation}\label{MSDx distributed 2}
\left\langle x^{2}(t)\right\rangle\simeq\frac{\log^{\nu/2}t}{\sqrt{\Gamma(1+\nu)}},
\end{equation}
and
\begin{equation}\label{MSDy distributed 2}
\left\langle y^{2}(t)\right\rangle\simeq2\frac{\log^{\nu}t}{\Gamma(1+\nu)},
\end{equation}
where we use the Tauberian theorem for slowly varying functions (see Appendix C). This is an ultraslow diffusion again. A prominent example of an ultraslow diffusion is the Sinai diffusion, where the MSD behaves as $\log^{4}t$ \cite{sinai,sinai_aljaz}. Logarithmically slow
diffusion also occurs in granular gases with constant restitution
coefficient \cite{anna}, for single file diffusion in disordered environments
\cite{lloyd}, or in dynamic maps \cite{kaerger}. Stochastic models leading to
logarithmically growing MSDs include continuous time random walks with superheavy
waiting time densities \cite{sinai_aljaz}, strongly localized heterogeneous
diffusion coefficients \cite{hdp_pccp}, ageing continuous time random walks
\cite{michael}, and ultraslow scaled Brownian motion \cite{anna1}.

\subsection{Cases with $\eta(t)=\frac{t^{-\alpha}}{\Gamma(1-\alpha)}$}

For power-law memory kernel $\eta(t)=\frac{t^{-\alpha}}{\Gamma(1-\alpha)}$ ($\eta(s)=s^{\alpha-1}$) and $\gamma(t)=\delta(t)$ it follows that $\xi(s)=s^{-\alpha+1/2}$. Since $\eta(t)$ is non-negative integrable function for $0<\alpha<1$, while $\xi(s)$ is completely monotone for $-\alpha+1/2<0$, and $s\xi(s)=s^{-\alpha+3/2}$ is Bernstein function for $0<-\alpha+3/2<1$, it follows that $1/2<\alpha<1$ should be satisfied in order $p_{1}(x,t)$ to be non-negative function. By using relation (\ref{p1 general}), Laplace transform formula (\ref{ML three Laplace}) yields
\begin{align}\label{ML power law}
p_{1}(k_{x},t)=E_{3/2-\alpha}\left(-\frac{1}{2}k_{x}^{2}t^{3/2-\alpha}\right),
\end{align}
from where by the Fourier transform formula (\ref{cosine H}), we find that the PDF $p_{1}(x,t)$ in terms of the Fox $H$-function (see definition (\ref{H_integral})) reads
\begin{align}\label{one power eta}
p_{1}(x,t)=\frac{1}{2|x|}H_{1,1}^{1,0}\left[\left.\frac{|x|}{\left(\frac{1}{2}t^{3/2-\alpha}\right)^{1/2}}\right|\begin{array}{c
l}
    \displaystyle {(1,3/4-\alpha/2)}\\
    \displaystyle {(1,1)}
  \end{array}\right].
\end{align}
For the MSD, relation (\ref{integral of H}) yields
\begin{align}\label{MSD one power eta}
\left\langle x^{2}(t)\right\rangle=\frac{t^{3/2-\alpha}}{\Gamma\left(5/2-\alpha\right)},
\end{align}
i.e. subdiffusive behavior for $1/2<\alpha<1$. The case $\alpha=1/2$ yields normal diffusion. From the asymptotic behavior of the Fox $H$-function \cite{saxena book}, we find that the PDF $p_{1}(x,t)$ has a stretched Gaussian form
\begin{align}\label{PDF one power eta asymptotic}
p_{1}(x,t)\simeq\frac{1}{\sqrt{2\pi(1+2\alpha)}}\frac{|x|^{\frac{1-2\alpha}{1+2\alpha}}}{\left(\frac{1}{2}t^{3/2-\alpha}\right)^{\frac{1}{1+2\alpha}}}
\exp\left(-\frac{1+2\alpha}{4}\left(\frac{3-2\alpha}{4}\right)^{\frac{3-2\alpha}{1+2\alpha}}\frac{|x|^{\frac{4}{1+2\alpha}}}{\left(\frac{1}{2}t^{3/2-\alpha}\right)^{\frac{2}{1+2\alpha}}}\right).
\end{align}
Note that the case $\alpha=1/2$ yields Gaussian form of PDF. For the $y$ direction, since $\gamma(t)=\delta(t)$, the diffusion is normal, i.e., $\left\langle y^{2}(t)\right\rangle=2t$.

In the case where $\gamma(t)=\eta(t)=\frac{t^{-\alpha}}{\Gamma(1-\alpha)}$, $0<\alpha<1$ (note that $\gamma(t)$ and $\eta(t)$ is non-negative integrable function), we recover Eq.(\ref{frac diffusion eq on a comb}), for which the PDF along the $x$ direction is given by
\begin{align}\label{p1 power power}
p_{1}(x,t)=\mathcal{F}^{-1}\left[E_{1-\alpha/2}\left(-\frac{1}{2}k_{x}^{2}t^{1-\alpha/2}\right)\right]=\frac{1}{2|x|}H_{1,1}^{1,0}\left[\left.\frac{|x|}{\left(\frac{1}{2}t^{1-\alpha/2}\right)^{1/2}}\right|\begin{array}{c
l}
    \displaystyle {(1,1/2-\alpha/4)}\\
    \displaystyle {(1,1)}
  \end{array}\right],
\end{align}
and the MSD becomes
\begin{align}\label{MSDy power power}
\left\langle x^{2}(t)\right\rangle=\frac{t^{1-\alpha/2}}{\Gamma\left(2-\alpha/2\right)}.
\end{align}
For the $y$ direction the results correspond to (\ref{p2 power delta}) and (\ref{MSDy power delta}). The non-negativity of $p_{1}(x,t)$ follows from the fact that $\xi(s)=s^{-\alpha/2}$ is completely monotone, and $s\xi(s)=s^{-\alpha/2+1}$ is a Bernstein function for $0<\alpha<1$. 

Two power-law memory function $\gamma(t)=C_{1}\frac{t^{-\alpha_{1}}}{\Gamma(1-\alpha_{1})}+C_{2}\frac{t^{-\alpha_{2}}}{\Gamma(1-\alpha_{2})}$, $0<\alpha_{1}<\alpha_{2}<1$, and power-law memory kernel $\eta(t)=\frac{t^{-\alpha}}{\Gamma(1-\alpha)}$, $0<\alpha<1$, the MSD along the $x$ direction becomes
\begin{align}\label{MSDy two power power}
\left\langle x^{2}(t)\right\rangle=\frac{1}{\sqrt{C_{2}}}t^{1+\alpha_{2}/2-\alpha} E_{\alpha_{2}-\alpha_{1},2+\alpha_{2}/2-\alpha}^{1/2}\left(-\frac{C_{1}}{C_{2}}t^{\alpha_{2}-\alpha_{1}}\right),
\end{align}
which in the short time limit behaves as $\left\langle x^{2}(t)\right\rangle\simeq\frac{1}{\sqrt{C_{2}}}\frac{t^{1+\alpha_{2}/2-\alpha}}{\Gamma\left(2+\alpha_{2}/2-\alpha\right)}$, and in the long time limit as $\left\langle x^{2}(t)\right\rangle\simeq\frac{1}{\sqrt{C_{1}}}\frac{t^{1+\alpha_{1}/2-\alpha}}{\Gamma\left(2+\alpha_{1}/2-\alpha\right)}$, i.e., decelerating anomalous subdiffusion. Here we note that in order PDF $p_{1}(x,t)$ to be non-negative, the following restrictions should be satisfied $0<\alpha_{1}/2<\alpha_{2}/2<\alpha<1$. This can be shown from the fact that $\xi(s)=\left(C_{1}s^{\alpha_1-2\alpha}+C_2s^{\alpha_2-2\alpha}\right)^{1/2}$ is a completely monotone function if $C_{1}s^{\alpha_1-2\alpha}$ and $C_{2}s^{\alpha_2-2\alpha}$ are completely monotone functions. This is satisfied if $\alpha_{1,2}-2\alpha<0$. Along the $y$ direction the MSD is given by (\ref{MSDy two power delta}) since the memory kernel $\gamma(t)$ is the same in both cases.

For the power-law memory kernel $\eta(t)=\frac{t^{-\alpha}}{\Gamma(1-\alpha)}$, $0<\alpha<1$, and uniformly distributed order $\gamma(t)$ (\ref{distibuted memory kernel gamma}) the MSD in the long time limit reads
\begin{equation}\label{MSDx distributed 1 eta}
\left\langle x^{2}(t)\right\rangle\simeq\frac{t^{1-\alpha}}{\Gamma(2-\alpha)}\log^{-1/2}t.
\end{equation}
This result is obtained with the use of Tauberian theorem (see Appendix C). 
The MSD along $y$ direction is the same as in Eq.(\ref{MSDy distributed 1}) since 
the same memory kernel $\gamma(t)$ is used.

\subsection{Cases with $\eta(t)=C_{1}\frac{t^{-\alpha_{1}}}{\Gamma(1-\alpha_{1})}+C_{2}\frac{t^{-\alpha_{2}}}{\Gamma(1-\alpha_{2})}$}

Furthermore, we consider kernel with two fractional exponents $\alpha_1<\alpha_2$, i.e., $\eta(t)=C_{1}\frac{t^{-\alpha_{1}}}{\Gamma(1-\alpha_{1})}+C_{2}\frac{t^{-\alpha_{2}}}{\Gamma(1-\alpha_{2})}$, $C_{1},C_{2}>0$, $C_{1}+C_{2}=1$, where $1/2\leq\alpha_{1}<\alpha_{2}<1$. This condition should be satisfied in order the corresponding PDF $p_{1}(x,t)$ to be non-negative and $\eta(t)$ to be non-negative integrable function. Let us show this. Since $\eta(t)$ should be non-negative integrable function then $0<\alpha_{1,2}<1$. From the kernels $\gamma(t)$ and $\eta(t)$ it follows that $\xi(s)=\frac{1}{C_{1}s^{\alpha_1-1/2}+C_{2}s^{\alpha_2-1/2}}$. From Appendix A, property 4, $\xi(s)$ is completely monotone if $C_{1}s^{\alpha_1-1/2}+C_{2}s^{\alpha_2-1/2}$ is complete Bernstein function. Thus, $0<\alpha_{1,2}-1/2<1$, i.e., $1/2<\alpha_{1,2}<3/2$. From here we obtain the restrictions on parameters $\alpha_{1,2}$, i.e., $1/2\leq\alpha_{1}<\alpha_{2}<1$. The PDF $p_{1}(x,t)$ is then obtained in terms of infinite series in the Fox $H$-functions,
\begin{align}\label{two powers eta}
p_{1}(x,t)&=\sum_{n=0}^{\infty}\frac{(-1)^{n}}{n!}\left(\frac{C_{1}}{C_{2}}\right)^{n}\nonumber\\&\times\frac{1}{2|x|} H_{3,3}^{2,1}\left[\left.\frac{|x|}{\left(\frac{1}{2}C_{2}t^{3/2-\alpha_{2}}\right)^{1/2}}\right|\left.
\begin{array}{c
l}
    \displaystyle {(1-n,1/2), ((\alpha_{2}-\alpha_{1})n+1,3-\alpha_{2}/2), (1,1/2)}\\
    \displaystyle {(1,1), (1,1/2), (1,1/2)}
  \end{array}\right.\right],
\end{align}
from where we find the MSD
\begin{align}\label{MSD two powers eta}
\left\langle x^{2}(t)\right\rangle=C_{1}\frac{t^{3/2-\alpha_{1}}}{\Gamma\left(5/2-\alpha_{1}\right)}
+C_{2}\frac{t^{3/2-\alpha_{2}}}{\Gamma\left(5/2-\alpha_{2}\right)}.
\end{align}
Therefore, for short time limit we obtain $\left\langle x^{2}(t)\right\rangle\simeq C_{2}\frac{t^{3/2-\alpha_{2}}}{\Gamma\left(5/2-\alpha_{2}\right)}$, whereas for long time limit  $\left\langle x^{2}(t)\right\rangle\simeq C_{1}\frac{t^{3/2-\alpha_{1}}}{\Gamma\left(5/2-\alpha_{1}\right)}$, i.e. accelerating anomalous diffusion. Generalization to the case of a kernel with many fractional exponents $\alpha_{i}$, $i=1,2,\dots,N$, is obvious. The diffusion along the $y$ direction is normal since $\gamma(t)=\delta(t)$.

In case of a power-law memory kernel $\gamma(t)=\frac{t^{-\alpha}}{\Gamma(1-\alpha)}$, such that $0<\alpha/2<\alpha_{1}<\alpha_{2}<1$ (these restrictions follow from the conditions (i) $\xi(s)=\frac{1}{C_{2}}\frac{s^{\alpha/2-\alpha_{1}}}{s^{\alpha_{2}-\alpha_{1}}+\frac{C_{1}}{C_{2}}}$ to be completely monotone, and (ii) $s\xi(s)$ to be the Bernstein function), for the MSD we find
\begin{align}\label{MSD power two powers}
\left\langle x^{2}(t)\right\rangle=C_{1}\frac{t^{1+\alpha/2-\alpha_{1}}}{\Gamma(2+\alpha/2-\alpha_{1})}+C_{2}\frac{t^{1+\alpha/2-\alpha_{2}}}{\Gamma(2+\alpha/2-\alpha_{2})},
\end{align}
from where for the short time limit MSD behaves as $\left\langle x^{2}(t)\right\rangle=C_{2}\frac{t^{1+\alpha/2-\alpha_{2}}}{\Gamma(2+\alpha/2-\alpha_{2})}$, and in the long time limit as $\left\langle x^{2}(t)\right\rangle=C_{1}\frac{t^{1+\alpha/2-\alpha_{1}}}{\Gamma(2+\alpha/2-\alpha_{1})}$. In the $y$ direction $\left\langle y^{2}(t)\right\rangle=2\frac{t^{\alpha}}{\Gamma(1+\alpha)}$.

In case where both memory kernels are combinations of two power-law functions $\gamma(t)=\eta(t)=C_{1}\frac{t^{-\alpha_{1}}}{\Gamma(1-\alpha_{1})}+C_{2}\frac{t^{-\alpha_{2}}}{\Gamma(1-\alpha_{2})}$, $0<\alpha_{1}<\alpha_{2}<1$, it follows $\xi(s)=\frac{\sqrt{\frac{1}{C_{2}}}s^{-\alpha_{1}/2}}{\left(s^{\alpha_{2}-\alpha_{1}}+\frac{C_{1}}{C_{2}}\right)^{1/2}}$. Thus, for the MSD along the $x$ direction we find
\begin{align}\label{MSD two powers two powers}
\left\langle x^{2}(t)\right\rangle=C_{2}^{1/2}\mathcal{L}^{-1}\left[\frac{s^{\alpha_{1}-2}}{\left(s^{\alpha_{2}-\alpha_{1}}+\frac{C_{1}}{C_{2}}\right)^{-1/2}}\right]=C_{2}^{1/2}t^{1-\alpha_{2}/2}E_{\alpha_{2}-\alpha_{1},2-\alpha_{2}/2}^{-1/2}\left(-\frac{C_{1}}{C_{2}}t^{\alpha_{2}-\alpha_{1}}\right),
\end{align}
from where the short time limit yields $\left\langle x^{2}(t)\right\rangle\simeq C_{2}^{1/2}\frac{t^{1-\alpha_{2}/2}}{\Gamma(2-\alpha_{2}/2)}$, and in the long time limit $\left\langle x^{2}(t)\right\rangle\simeq C_{1}^{1/2}\frac{t^{1-\alpha_{1}/2}}{\Gamma(2-\alpha_{1}/2)}$. The MSD in $y$ direction is given by Eq.(\ref{MSDy two power delta}).

A more complicated combination follows if $\gamma(t)$ is uniformly distributed memory kernel (\ref{distibuted memory kernel gamma}), for which $\xi(s)=\frac{1}{C_{2}}\frac{s^{-\alpha_{1}}\sqrt{\frac{s-1}{\log{s}}}}{s^{\alpha_{2}-\alpha_{1}}+\frac{C_{1}}{C_{2}}}$. For the MSD in $x$ direction, by using Tauberian theorem (see Appendix C), we derive
\begin{align}\label{MSD distributed two powers}
\left\langle x^{2}(t)\right\rangle\simeq C_{1}\frac{t^{1-\alpha_{1}}}{\Gamma(2-\alpha_{1})}\log^{-1/2}{t}+C_{2}\frac{t^{1-\alpha_{2}}}{\Gamma(2-\alpha_{2})}\log^{-1/2}{t}.
\end{align}
From here the previously obtained result (\ref{MSDx distributed 1 eta}) follows directly.
In $y$ direction the MSD is given by Eq.(\ref{MSDy distributed 1}).

\subsection{Cases with distributed order kernel $\eta(t)$}

The case where $\gamma(t)=\delta(t)$ and $\eta(t)$ is a uniformly distributed order memory kernel, i.e., its Laplace transform is given by $\eta(s)=\frac{s-1}{s\log s}$, can not be considered
within our framework, since $\xi(s)\sim\frac{s^{1/2}\log s}{s-1}$ is not a completely monotone function. 

Similar situation appears for power-law memory kernel $\gamma(t)$ (and two-power law memory kernels $\gamma(t)$) and uniformly distributed memory kernel $\eta(t)$, for which $\xi(s)\sim\frac{s^{\alpha/2}\log s}{s-1}$ is not a completely monotone function.

For the case where both kernels are of distributed order, i.e., $\gamma(t)=\eta(t)=\int_{0}^{1}d\alpha\,\frac{t^{-\alpha}}{\Gamma(1-\alpha)}$, we find
in the long time limit with the use of Tauberian theorem (see Appendix C),
\begin{equation}\label{MSDx distributed 1 1}
\left\langle x^{2}(t)\right\rangle\simeq t\log^{-1/2}t.
\end{equation}
The MSD along $y$ direction is the same as in Eq.(\ref{MSDy distributed 1}).

\section{Summary}

In this paper we studied anomalous diffusion in a generalized two-dimensional comb-like 
model governed by diffusion-like equation with two memory kernels.
We solved this equation in the Fourier-Laplace space and  gave general expressions 
for the PDFs and MSDs in $x$ and $y$ directions. The CTRW model, which corresponds to the considered diffusion-like equation on a comb-like structure was presented. We showed that the waiting time PDF which is responsible for the appearance of different diffusive behavior along the backbone ($x$ direction) depends on both memory kernels $\gamma(t)$ and $\eta(t)$, whereas the waiting time PDF along $y$ direction depends only on the memory kernel $\gamma(t)$. We found the PDFs which subordinate the random diffusion processes on the comb-like structure considered to the Wiener process. We investigated the role of different forms of $\gamma(t)$ and $\eta(t)$, such as Dirac delta, power-law, and distributed order. The results obtained for the PDFs and MSDs are represented by using the Mittag-Leffler and Fox $H$-functions, from where different diffusive regimes can be observed. It is shown that the considered model may be used to describe anomalous diffusive processes, including decelerating and accelerating anomalous subdiffusion, and ultraslow diffusion as well. The results for the MSDs in $x$ and $y$ directions are summarized in Tables I and II.

\begin{table}[h]\label{tab1}
\centering{\caption{MSD $\left\langle x^{2}(t)\right\rangle$ along $x$ direction. It depends on both memory kernels $\gamma(t)$ and $\eta(t)$. The MSDs in case of distributed order memory kernels are calculated in the long time limit by using Tauberian theorem.}}
\begin{tabular}{|l|l|l|l|l|}
\hline
     $\eta(t) \quad \setminus \quad \gamma(t)$ &   $\delta(t)$       & $\frac{t^{-\alpha}}{\Gamma(1-\alpha)}$     & 
     $         C_{1}\frac{t^{-\alpha_{1}}}{\Gamma(1-\alpha_{1})}
         +C_{2}\frac{t^{-\alpha_{2}}}{\Gamma(1-\alpha_{2})}$ & $\int_{0}^{1}d\alpha\,\frac{t^{-\alpha}}{\Gamma(1-\alpha)}$     \\ \hline
  $\delta(t)$ & $\sim t^{1/2}$       & $\sim t^{\alpha/2}$ & $\sim t^{\alpha_{2}/2}E_{\alpha_{2}-\alpha_{1},\alpha_{2}/2+1}^{1/2}\left(-\frac{C_{1}}{C_{2}}t^{\alpha_{2}-\alpha_{1}}\right)$ & $\sim \log^{1/2}t$
  \\ \hline
  $\frac{t^{-\alpha}}{\Gamma(1-\alpha)}$ & $\sim t^{3/2-\alpha}$     & $\sim t^{1-\alpha/2}$ & $\sim t^{1+\alpha_{2}/2-\alpha}E_{\alpha_{2}-\alpha_{1},2+\alpha_{2}/2-\alpha}^{1/2}\left(-\frac{C_{1}}{C_{2}}t^{\alpha_{2}-\alpha_{1}}\right)$ & $\sim t^{1-\alpha}\log^{-1/2}t$
  \\ \hline
  $\begin{array}{c
       l}
           C_{1}\frac{t^{-\alpha_{1}}}{\Gamma(1-\alpha_{1})}\\
           +C_{2}\frac{t^{-\alpha_{2}}}{\Gamma(1-\alpha_{2})}
         \end{array}$
  & $\begin{array}{c
         l}
             \sim C_{1}\frac{t^{3/2-\alpha_{1}}}{\Gamma\left(5/2-\alpha_{1}\right)}\\
             +C_{2}\frac{t^{3/2-\alpha_{2}}}{\Gamma\left(5/2-\alpha_{2}\right)}
           \end{array}$  &
 $\begin{array}{c
          l}
              \sim C_{1}\frac{t^{1+\alpha/2-\alpha_{1}}}{\Gamma\left(2+\alpha/2-\alpha_{1}\right)}\\
              +C_{2}\frac{t^{1+\alpha/2-\alpha_{2}}}{\Gamma\left(2+\alpha/2-\alpha_{2}\right)}
            \end{array}$ &
 $\sim t^{1-\alpha_{2}/2}E_{\alpha_{2}-\alpha_{1},2-\alpha_{2}/2}^{-1/2}\left(-\frac{C_{1}}{C_{2}}t^{\alpha_{2}-\alpha_{1}}\right)$ &
 $\begin{array}{c
           l}
               \sim C_{1}\frac{t^{1-\alpha_{1}}}{\Gamma\left(2-\alpha_{1}\right)}\log^{-1/2}t\\
               +C_{2}\frac{t^{1-\alpha_{2}}}{\Gamma\left(2-\alpha_{2}\right)}\log^{-1/2}t
             \end{array}$    \\ \hline
  $\int_{0}^{1}d\alpha\,\frac{t^{-\alpha}}{\Gamma(1-\alpha)}$ & $-$     & $-$  & $-$ & $\sim t\log^{-1/2}t$ \\ \hline
\end{tabular}
\end{table}

\begin{table}[h]\label{tab2}
\centering{\caption{MSD $\left\langle y^{2}(t)\right\rangle$ along $y$ direction. 
It depends only on the memory kernel $\gamma(t)$.}}
\begin{tabular}{|l|l|l|l|l|}
\hline
     $\gamma(t)$       & $\delta(t)$       & $\frac{t^{-\alpha}}{\Gamma(1-\alpha)}$     & 
     $         C_{1}\frac{t^{-\alpha_{1}}}{\Gamma(1-\alpha_{1})}
         +C_{2}\frac{t^{-\alpha_{2}}}{\Gamma(1-\alpha_{2})}$ & $\int_{0}^{1}d\alpha\,\frac{t^{-\alpha}}{\Gamma(1-\alpha)}$     \\ \hline
  $\left\langle y^{2}(t)\right\rangle$       & $\sim t$       & $\sim t^{\alpha}$ & $\sim t^{\alpha_{2}}E_{\alpha_{2}-\alpha_{1},\alpha_{2}+1}\left(-\frac{C_{1}}{C_{2}}t^{\alpha_{2}-\alpha_{1}}\right)$ & $\sim \gamma+\log{t}+e^{t}\mathrm{E}_{1}(t)$ \\ \hline
  \end{tabular}
\end{table}

\section*{Acknowledgments}
AI thanks the support  from the Israel Science Foundation (ISF-1028). RM acknowledges
financial support from the Academy of Finland within the Finland Distinguished
Professor programme. TS and AC acknowledge the hospitality and support from the MPIPKS. 

\appendix

\section{Completely monotone and Bernstein functions}

Here we give definitions and some properties of completely monotone and Bernstein functions \cite{book bernstein}.

\begin{enumerate}
\item The function $g(s)$ is a completely monotone if $(-1)^{n}g^{(n)}(s)\geq0$ for all $n\geq0$ and $s>0$. Product of completely monotone functions is completely monotone function too. An example of completely monotone function is $s^{\alpha}$, where $\alpha<0$.

\item The Bernstein characterization theorem states that if the Laplace transform $g(s)=\mathcal{L}\left[G(t)\right]$ of given function $G(t)$ is completely monotone function, then the function $G(t)$ is non-negative.

\item The function $e^{-us\xi(s)}$ is a completely monotone if $s\xi(s)$ is a complete Bernstein function. A given function $f(s)$ is a Bernstein function if $(-1)^{n-1}f^{(n)}(s)\geq0$ for all $n\in N$ and $s>0$. An example of Bernstein function is $s^{\alpha}$, where $0<\alpha<1$.

\item It can be shown that if $f(s)$ is a complete Bernstein function, then $g(s)=1/f(s)$ is a completely monotonic function \cite{berg}.

\item Another important property of complete Bernstein function is the one which states that $f(s)$ is a complete Bernstein function if and only if the function $s/f(s)$ is
complete Bernstein function. 
\end{enumerate}
In order $h_{1}(u,s)$ (\ref{h1(u,s)}) to be completely monotone function, both functions $\xi(s)=\frac{1}{\eta(s)}\sqrt{\frac{\gamma(s)}{s}}$ and $e^{-us\xi(s)}$ should be completely monotone \cite{book bernstein}. The function $e^{-us\xi(s)}$ is a completely monotone if $s\xi(s)=\frac{1}{\eta(s)}\sqrt{s\gamma(s)}$ is a Bernstein function.  Similarly, in order the PDF $h_{2}(u,s)$ to be completely monotone function, the function $\gamma(s)$ should be completely monotone and  $s\gamma(s)$ is a Bernstein function.

\section{Mittag-Leffler and Fox $H$-functions}

To calculate the PDFs we use the three parameter Mittag-Leffler (M-L) function \cite{Prabhakar}
\begin{equation}\label{ML three}
E_{\alpha,\beta}^{\delta}(z)=\sum_{k=0}^{\infty}\frac{(\delta)_k}{\Gamma(\alpha
k+\beta)}\frac{z^k}{k!},
\end{equation}
where $(\delta)_k=\Gamma(\delta+k)/\Gamma(\delta)$ is the Pochhammer symbol, which Laplace transform is given by \cite{Prabhakar}
\begin{equation}\label{ML three Laplace}
\mathcal{L}\left[t^{\beta-1}E_{\alpha,\beta}^{\delta}(\pm
at^{\alpha})\right](s)=\frac{s^{\alpha\delta-\beta}}{\left(s^{\alpha}\mp
a\right)^{\delta}}, \quad \Re(s)>|a|^{1/\alpha}.
\end{equation}
The case $\delta=1$ yields the two parameter M-L function $E_{\alpha,\beta}(z)$, and the case $\beta=\delta=1$ - one parameter M-L function $E_{\alpha}(z)$. For the three parameter M-L function the following formula holds \cite{Saxena,sandev pla,sandev jmp} (see also \cite{seybold,huang} for two parameter M-L function)
\begin{equation}\label{ML three asymptotic}
E_{\alpha,\beta}^{\delta}(-z)=\frac{z^{-\delta}}{\Gamma(\delta)}\sum_{n=0}^{\infty}\frac{\Gamma(\delta+n)}{\Gamma(\beta-\alpha(\delta+n)
)}\frac{(-z)^{-n}}{n!}, \quad |z|>1,
\end{equation}
in order to analyze the asymptotic behaviors. Thus, the asymptotic expansion formula for three parameter M-L function is given by
$E_{\alpha,\beta}^{\delta}(-z) \simeq\frac{z^{-\delta}}{\Gamma(\beta-\alpha\delta)}$
for $z\rightarrow\infty$.

The Fox $H$-function (or simply $H$-function) is defined by the
following Mellin-Barnes integral \cite{saxena book}
\begin{align}\label{H_integral}
H_{p,q}^{m,n}\left[z\left|\begin{array}{c l}
    (a_1,A_1),...,(a_p,A_p)\\
    (b_1,B_1),...,(b_q,B_q)
  \end{array}\right.\right]=H_{p,q}^{m,n}\left[z\left|\begin{array}{c l}
    (a_p,A_p)\\
    (b_q,B_q)
  \end{array}\right.\right]=\frac{1}{2\pi\imath}\int_{\Omega}\theta(s)z^{s}{d}s,
\end{align}
where
$\theta(s)=\frac{\prod_{j=1}^{m}\Gamma(b_j-B_js)\prod_{j=1}^{n}\Gamma(1-a_j+A_js)}{\prod_{j=m+1}^{q}\Gamma(1-b_j+B_js)\prod_{j=n+1}^{p}\Gamma(a_j-A_js)}$,
$0\leq n\leq p$, $1\leq m\leq q$, $a_i,b_j \in C$, $A_i,B_j \in
R^{+}$, $i=1,...,p$, $j=1,...,q$. The contour $\Omega$ starting at
$c-i\infty$ and ending at $c+i\infty$ separates the poles
of the function $\Gamma(b_j+B_js)$, $j=1,...,m$ from those of the
function $\Gamma(1-a_i-A_is)$, $i=1,...,n$.

The connection between three parameter M-L function and Fox $H$-function is given by \cite{saxena book}
\begin{eqnarray}\label{HML}
E_{\alpha,\beta}^{\delta}(-z)=\frac{1}{\Gamma(\delta)}H_{1,2}^{1,1}\left[z\left|\begin{array}{l}
    (1-\delta,1)\\
    (0,1),(1-\beta,\alpha)
  \end{array}\right.\right].
\end{eqnarray}

The Fourier transform formula for Fox $H$-function is \cite{saxena book}
\begin{align}\label{cosine H}
\int_{0}^{\infty}dk\,k^{\rho-1}\cos(kx)H_{p,q}^{m,n}\left[ak^{\delta}\left|\begin{array}{c
l}
    (a_p,A_p)\\
    (b_q,B_q)
  \end{array}\right.\right]=\frac{\pi}{x^\rho}H_{q+1,p+2}^{n+1,m}\left[\frac{x^\delta}{a}\left|\begin{array}{c
l}
     (1-b_q,B_q), (\frac{1+\rho}{2}, \frac{\delta}{2})\\
    (\rho,\delta), (1-a_p,A_p), (\frac{1+\rho}{2},\frac{\delta}{2})
  \end{array}\right.\right],\nonumber\\
\end{align}
where $\Re\left(\rho+\delta \min_{1\leq j\leq
m}\left(\frac{b_j}{B_j}\right)\right)>1$, $x^\delta>0$,
$\Re\left(\rho+\delta \max_{1\leq j\leq
n}\left(\frac{a_j-1}{A_j}\right)\right)<\frac{3}{2}$,
$|\arg(a)|<\pi\theta/2$, $\theta>0$,
$\theta=\sum_{j=1}^{n}A_j-\sum_{j=n+1}^{p}A_j+\sum_{j=1}^{m}B_j-\sum_{j=m+1}^{q}B_j$.

The Mellin transform of the $H$-function is given by \cite{saxena book}
\begin{eqnarray}\label{integral of H}
&\int_{0}^{\infty}x^{\xi-1}H_{p,q}^{m,n}\left[ax\left|\begin{array}{c
l}
    (a_p,A_p)\\
    (b_q,B_q)
  \end{array}\right.\right]{d}x=a^{-\xi}\theta(-\xi),\nonumber\\
\end{eqnarray}
where $\theta(-\xi)$ is defined in Eq.(\ref{H_integral}).

\section{Tauberian theorem for slowly varying functions \cite{feller}}

If some function $r(t)$, $t\geq 0$, has the Laplace transform $r(s)$ whose asymptotics behaves as
\begin{equation}\label{tauber7}
r(s)\simeq s^{-\rho}L\left(\frac{1}{s}\right), \quad
s\rightarrow0, \quad \rho\geq0,
\end{equation}
then
\begin{equation}\label{tauber8}
r(t)=\mathcal{L}^{-1}\left[r(s)\right]\simeq
\frac{1}{\Gamma(\rho)}t^{\rho-1}L(t), \quad t\rightarrow\infty.
\end{equation}
Here $L(t)$ is a slowly varying function  at infinity, i.e., $\lim_{t\rightarrow\infty}\frac{L(at)}{L(t)}=1$, for any $a>0$.

\end{document}